\documentclass{emulateapj}
\received{2007 October 9}
\accepted{2008 March 2}
\slugcomment{}
\shorttitle{MgII Line Variability of High Luminosity Quasars}
\shortauthors{Woo}

\newcommand{\kms}{${\rm kms}^{-1}$}

\begin{document}

\title{MgII Line Variability of High Luminosity Quasars}

\author{Jong-Hak Woo}
%\author{Jong-Hak Woo\altaffiltext{1,2}}
\affil{Department of Physics, University of California Santa Barbara, CA 93106-9530}
\affil{Yale Center for Astronomy and Astrophysics, Yale University, New Haven, CT 06520-8101}
%\altaffiltext{1}{Department of Physics, University of California, Santa Barbara, CA 93106-9530; woo@physics.ucsb.edu.  } 
\email{woo@physics.ucsb.edu}

\begin{abstract}
We monitored five high-luminosity quasars with $\lambda L_{3000\AA} > 10^{45}$ erg s$^{-1}$ at $0.4 \lesssim z \lesssim 0.6$ 
to measure flux variability of the MgII $\lambda$2798 line and explore 
feasibility of reverberation mapping using MgII.
Over the two year monitoring program,
imaging data were obtained with the A Noble Double-Imaging Camera 
on the 1.3-m telescope at the Cerro Tololo Inter-American Observatory (CTIO),
while spectroscopic data were obtained at the same night 
with the R-C spectrograph on the 1.5-m telescope at the CTIO. 
By performing differential photometry using available field stars in each quasar image,
we measured variability -- 10\%-24\% peak-to-peak changes and 3\%-8\% 
rms variations -- in the B band, which includes flux changes in the
rest-frame UV continuum ($\sim$2500\AA -- $\sim$3600\AA) as well as the MgII line.
Utilizing photometric measurements for spectroscopic flux calibration,
we measured the MgII line flux and the continuum flux at 3000\AA~ 
from each single-epoch spectrum.
Four objects showed MgII line flux variability with
23\%-50\% peak-to-peak changes and  8\%-17\% rms variations
over 1-1.5 year rest-frame time scales,
while one object showed no MgII flux variability within the measurement
error ($<$5\%). We also detected 4\%-15\% rms variations of the MgII
line width for all five objects.
With synchronous observations for photometry and spectroscopy, 
we demonstrated the feasibility of the MgII line reverberation mapping 
for high-luminosity quasars at intermediate redshift.
\end{abstract}


\section{Introduction}

Determining black hole mass is a fundamental step to understand
the physics of the AGN accretion disk (e.g. Bonning et al. 2007; Davis et al. 2007)
%as well as the role of black holes in galaxy formation and evolution
%(e.g. Di Matteo et al. 2005; Croton et al. 2006) and cosmic evolution
as well as cosmic evolution of black holes and galaxies
(e.g. Di Matteo et al. 2005; Croton et al. 2006; Woo et al. 2006; Treu et al. 2007).
Active galaxies with broad emission lines provide a way of 
estimating black hole masses, assuming
that the motion of the broad-line regions (BLR) is dominated
by the gravitational potential of the central source. Thus,
virial mass can be determined from the dynamics of the BLR. 
The width of broad-emission lines are used in estimating the velocity scale,
while the size (R$_{\rm BLR}$) can be determined by measuring
the time lag between the continuum flux changes (presumably
originating from an accretion flow) and the corresponding flux changes in
the broad-emission lines (Blanford \& McKee 1982).

So far, reverberation masses have been determined for about three dozens 
of local Seyfert galaxies and PG quasars, which are limited to redshift $z<0.4$ and
optical luminosity L$_{5100\AA}$ $<$ 10$^{46}$ erg s$^{-1}$
(Wandel et al.\ 1999; Kaspi et al. 2000, Peterson et al. 2004), 
while there are 
numerous AGNs with higher optical luminosity
%up to 10$^{48}$ erg s$^{-1}$
at $z>0.4$. 
Thus, it is crucial to increase the sample size, especially
including higher luminosity and higher redshift AGNs, in order to
validate the virial technique and understand its uncertainties.
The size-luminosity relation -- correlation between the measured
time lag (BLR size) and the continuum luminosity (Kaspi et al. 2005;
Bentz et al. 2006) -- provides a simple recipe for black hole mass estimation 
without reverberation mapping 
(e.g. Woo \& Urry 2002; Vestergaard 2002; Netzer \& Trakhtenbrot 2007; McGill et al. 2007).
Therefore, measuring time lags for higher luminosity AGNs is also necessary to
validate the size-luminosity relation and avoid extrapolations in estimating
black hole masses for high luminosity and high redshift quasars
(e.g. McLure \& Dunlop 2004; Vestergaard 2004).

Depending on the redshift of the source, different broad emission lines can be
used for reverberation studies.
For local AGNs, the best studied Balmer lines, such as H$\beta$ and H$\alpha$, 
showed clear correspondence to the continuum changes (e.g. Peterson et al. 2004).
At $z > 2$, CIV $\lambda$1549 and Ly$\alpha$ lines are redshifted in 
the optical range and 
become useful for reverberation studies with optical spectrographs. 
Recently, Kaspi et al. (2007) showed feasibility of 
reverberation mapping for quasars at $z > 2$,
reporting CIV line flux variability for six quasars and
a tentative time lag measurement for one object. 
In the same study, however, the Ly$\alpha$ line showed 
no flux variability within 5\% measurement error.
In the local universe, time lag measurements with CIV line have been 
reported for several Seyfert galaxies (see Peterson et al. 2004, for a summary).

At intermediate redshift, $0.7 < z < 2$, the MgII $\lambda$2798 line becomes virtually 
the only available broad line for estimating virial black hole masses.
However, there has been only a handful of studies on MgII line flux variability, 
and it is not clear whether the MgII line shows the same reverberation as other broad lines.
%Several groups reported that  MgII line flux variability of local AGNs is weaker than other broad lines. 
For example, Clavel et al. (1991) reported that MgII line flux variability of NGC 5548 was 
much lower than flux changes in other broad lines and did not show clear correspondence to UV continuum flux changes.
In the case of NGC 3516, the MgII line flux was almost constant in 1996 
although other lines and the continuum showed large flux variations (Goad et al. 1999). 
In contrast, time-lag measurements were consistent between MgII and
Balmer lines for NGC 3783 (Reichert et al. 1994) 
and NGC 4151 (Metzroth et al. 2006).
At high redshift, $1<z<2$, Trevese et al. (2007) studied  two high-luminosity quasars and detected 
MgII line flux variability for one object.
Clearly, flux variability of the MgII line has to be better constrained
to test feasibility of reverberation mapping,
especially for intermediate-to-high redshift quasars.

We initiated a monitoring program to study MgII line variability and 
explore feasibility of reverberation mapping for bright quasars
with the 1.5-m and 1.3-m 
telescopes at the Cerro Tololo Inter-American Observatory (CTIO)
when these telescopes became available in service mode        
through the Small and Moderate Aperture Research Telescope System (SMARTS) consortium\footnote{http://www.astro.yale.edu/smarts} in 2003.
In this paper we report our imaging and spectroscopic study on the MgII line variability of 
five high-luminosity quasars at $0.4 < z < 0.6$.
The goal of the paper is to investigate variability of the MgII line flux and width.
%to explore the feasibility of reverberation mapping with the MgII line.
%; ii) study variability of the MgII line width and line profile
%to determine the contribution to the systematic uncertainty of 
%the black hole mass estimated from single epoch spectra.
The paper is organized as follows. In \S~2 we describe observations and data
reduction. 
In \S~3 we describe our measurements of B band luminosity variability 
and MgII line flux variability.
%In \S~4 we investigate line width variation and uncertainties of black hole mass estimates
%based on the MgII line width. 
In \S~4 we discuss our results and their implications. A standard
cosmology is assumed where necessary (H$_0$=70 \kms
Mpc$^{-1}$, $\Omega_{\rm m}=0.3$ and $\Omega_{\Lambda}=0.7$).

\section{Observations and Data Reduction}

\begin{deluxetable*}{lcrrrrr}
%\rotate {}
\tablewidth{0pt}
\tablecaption{Target properties}
\tablehead{
\colhead{(1) Name}        &
\colhead{(2) z}           &
\colhead{(3) RA (J2000)}  &
\colhead{(4) DEC (J2000)} &
\colhead{(5) B}            &    
\colhead{(6) log M$_{BH}$}  &  
\colhead{(7) L/L$_{Edd}$}    } 
\tablecolumns{7}
\startdata
CTS300 & 0.352      &  11 28 18.62 & -13 19 28.7 & 17.2 & 8.74 & 0.12\\
CTS306 & 0.456      &  11 59 41.02 & -19 59 24.7 & 17.2 & 8.87 & 0.13\\
CTS536 & 0.500      &  14 01 49.27 & -24 55 30.3 & 16.5 & 8.81 & 0.40 \\
CTS582 & 0.483      &  22 40 18.84 & -52 31 52.4 & 17.8 & 8.75 & 0.13 \\
CTS667 & 0.557      &  11 39 10.70 & -13 50 43.6 & 16.7 & 8.62 & 1.12 \\
\enddata
\label{T_target}
\tablecomments{
Col. (1): Target ID.  
Col. (2): Redshift measured from the MgII line.
Col. (3): RA.
Col. (4): DEC.
Col. (5): mean B magnitude.
Col. (6): black hole mass estimated using the MgII FWHM and
the continuum luminosity at 3000$\AA$ from the mean spectra.
Col. (7): Eddington ratio calculated using black hole mass and 5.9$\times$ L$_{3000\AA}$ as bolometric luminosity.
}
\end{deluxetable*}

\begin{deluxetable*}{lcr}
\tablewidth{0pt}
\tablecaption{Journal of observations}
\tablehead{
\colhead{Date}         &
\colhead{Photomety}    &
\colhead{Spectroscopy} \\
\colhead{}              &
\colhead{Targets}       &
\colhead{Targets}       \\
\colhead{(1)} &
\colhead{(2)} &
\colhead{(3)}}  
\tablecolumns{3}
\startdata
2003 Mar 12 &  CTS306, CTS536   & CTS306, CTS536 \\
2003 Apr 4  &                   & CTS306$^{a}$, CTS536$^{a}$ \\
2003 Apr 5  &  CTS306, CTS536   &                \\
2003 May 7  &  CTS306, CTS536   & CTS306, CTS536 \\
2003 Jun 17 &  CTS582           & CTS536$^{a}$, CTS582 \\
2003 Jun 18 &  CTS306, CTS536   &                \\
2003 Jun 19 &  CTS306, CTS536   &                \\
2003 Jun 20 &  CTS306, CTS536, CTS582  &         \\
2003 Jun 25 &  CTS306                  & CTS306  \\
2003 Jul 20 &  CTS306, CTS536, CTS582  & CTS306, CTS536 \\
2003 Sep 23 &                   & CTS582$^{a}$ \\
2003 Oct 21 &  CTS582           & CTS582 \\
2003 Nov 12 &  CTS582           & CTS582 \\
2003 Dec 16 &  CTS582           & CTS582 \\
2003 Dec 19 &  CTS300           &        \\ 
2003 Dec 20 &  CTS667           &        \\
2003 Dec 29 &                   & CTS300$^{b}$  \\
2003 Dec 30 &                   & CTS667$^{b}$ \\
2004 Jan 20 &  CTS300, CTS667   & CTS667 \\
2004 Jan 21 &  CTS306, CTS536   & CTS306, CTS536 \\
2004 Feb 23 &  CTS536           & CTS536 \\
2004 Mar 2  &  CTS300           & CTS300 \\
2004 Mar 16 &  CTS667           & CTS667 \\
2004 Mar 31 &  CTS300, CTS536   & CTS300, CTS536 \\
2004 Apr 15 &  CTS300, CTS306, CTS536, CTS667 &  CTS300, CTS306, CTS536, CTS667 \\              
2004 Apr 16 &  CTS582           & CTS582 \\
2004 May 13 &  CTS582           &        \\               
2004 Jun 22 &  CTS536           & CTS536, CTS582$^{b}$  \\
2004 Jul 14 &  CTS536           & CTS536  \\
2004 Aug 11 &  CTS582            & \\
2004 Sep 9  &  CTS582           & \\
2004 Nov 16 &                   & CTS582$^{b}$  \\
2004 Dec 18 &  CTS582           & CTS582 \\
2004 Dec 20 &  CTS667           & CTS667 \\
2004 Dec 21 &  CTS306           & CTS306 \\
2004 Dec 22 &                   & CTS300$^{b}$ \\
2005 Jan 12 &  CTS667           & CTS667 \\
2005 Jan 13 &  CTS306           &        \\
2005 Jan 14 &  CTS300           & CTS300  \\
2005 Jan 15 &                   & CTS306$^{b}$  \\
2005 Feb 8  &                   & CTS300$^{b}$  \\
2005 Feb 9  &                   & CTS667$^{b}$  \\
\enddata
\label{T_journal}
\tablecomments{
Col. (1): Observing date.
Col. (2): Photometry targets
Col. (3): Spectroscopy targets
}
\tablerefs{
a) these spectra were flux calibrated with photometric measurements from the previous or following night.
b) these spectra were excluded in the analysis of MgII line flux variability due to the lack of photometric data 
}
\end{deluxetable*}
\subsection{Sample Selection and Experimental Design}

To study feasibility of reverberation mapping with the MgII line, 
a sample of bright southern quasars was selected from the 
Calan-Tololo Survey catalogue (Maza et al. 1993, 1995, 1996) 
with a redshift range, $0.4 \lesssim z \lesssim 1$, where no reverberation study has been reported.
Twelve bright objects were initially selected based on the observability 
of more than 6 months per year. 
Here we present five best observed objects with high quality spectra and strong MgII lines,
observed at more than 5 different epochs for spectroscopy and photometry.
Target properties of these objects are listed in Table 1.
The other seven objects were excluded in this study due to the fact 
that observed spectra have lower signal-to-noise ratio 
($\lesssim 10$), not enough to measure line flux variations with $\sim$5\% measurement errors, and$/$or the number of observed epochs with both photometry and spectroscopy
is typically less than 5. 

We designed spectroscopic and photometric observations to be carried 
out at the same night so that spectroscopic flux calibration can be 
accurately obtained using photometric measurements,
assuming that there is no intra-night flux variations (see \S 3.2).
However, because of the difficulty of scheduling,
there are several epochs when spectroscopy and photometry were 
carried out at different nights.  
We excluded data from those epochs for the analysis of the MgII line flux variations (see Table 2).

\begin{figure}
\epsscale{1.3}
\plotone{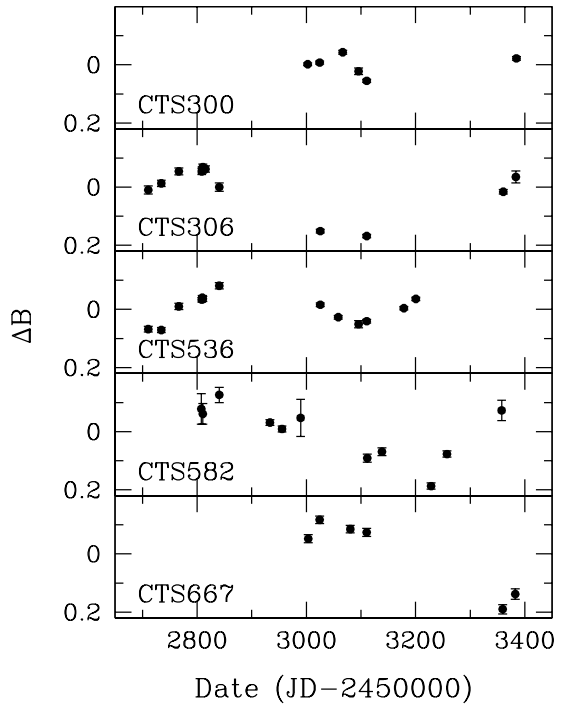}
\caption{B band (rest-frame $\sim$2500\AA-$\sim$3600\AA) light curves of all five objects.
Peak-to-peak variations are 9\%-24\% and rms variations are 3\%-8\%,
calculated in logarithmic scale.
}
\end{figure}

\subsection{Photometry}

Photometric and spectroscopic observations were performed in service mode,
respectively, with the CTIO 1.3-m and 1.5-m telescopes operated by the SMARTS consortium  
between March 2003 and February 2005. 
Due to the observability and limited telescope time,
$\sim$10 single-epoch data were obtained for given objects during the monitoring program.
Table 2 shows journal of observations.

For photometry, we used the A Noble Double-Imaging Camera (ANDICAM), a dual-channel optical
IR imager at the CTIO 1.3-m telescope. The optical CCD with an excellent cosmetics 
provided $\sim 6\times 6$ arcminute$^{2}$ field of view and 0.369'' per pixel 
resolution. Each target quasar was roughly centered in the field of view 
to cover as many as possible field stars for differential photometry.
We used the broad B band filter to match the observed wavelength range in spectroscopy,
in order to do spectroscopic flux calibration using photometric measurements.
The typical exposure time was 10 minute, yielding signal-to-noise ratio (S/N) $>$ 100 
on the target quasars. Bias and dome flats were taken in the afternoon before
each night. 
A set of standard stars were also taken during the night.
Initial data reduction, including bias subtraction and flat-fielding,
has been automatically performed through a pipeline
by the Yale SMARTS team. 

\subsection{Spectroscopy}

We used the R-C spectrograph with the 600 line mm$^{-1}$ grating
centered at $\sim$ 4500\AA. The observed spectral range was $\sim$3500-5300\AA, 
covering the MgII line (rest wavelength $\sim 2798$\AA) near the center in the
observed spectral range. 
The pixel scale corresponded to 1.47 \AA~ and the spatial scale along the slit 
was 1$\farcs$3 per pixel. 
We chose the 2$''$ wide slit to include most of the AGN light within the slit 
under the typical sub-arsecond seeing conditions at the CTIO. The spectral resolution
measured from arc lines was $\sim$ 3.8\AA~(FWHM), yielding Gaussian velocity of 
110 km s$^{-1}$ at the center of the spectral range.  
Total exposure time for each object was 3600 second, divided into 3
exposures in order to remove cosmic rays, 
yielding S/N $\sim$ 10-30 per pixel on the continuum.
Internal lamp flats were taken in the afternoon to correct pixel-to-pixel variations. 
A set of spectrophotometric standards were observed during each night
for initial flux calibrations (see \S 3.2). 

We performed the standard data reduction including bias subtraction,
flat fielding, wavelength calibration, spectral extraction, and flux calibration
using a series of IRAF scripts.  
He and Ar arc lines were used for wavelength calibration.
One-dimensional spectra were extracted for maximal S/N with a typical extraction
radius of 3 pixels, corresponding to $\sim$ 3.9$''$.
Individual spectra were combined with inverse-variance weighting to create the
final spectrum for each single-epoch.
For each night, we obtained sensitivity function using spectrophotometric stars. 
The flux scale is consistent within 1\% rms variation over the monitoring period.

\section{Measurements}

In this section, we present broad-band flux variability measured from 
differential photometry (\S 3.1). 
Then we show our spectroscopic analysis for MgII line variability.
First, we demonstrate spectroscopic flux calibration with photometric measurements (\S 3.2). 
In \S 3.3 we present how we subtract the continuum and measure the MgII line flux. 
We report MgII line flux variability in \S 3.4, and MgII line width variability
in \S 3.5.

\subsection{B band flux variability from photometry}

To measure the broad band flux variability, 
we performed differential photometry using stars in the field of each quasar,
as demonstrated in the paper by Woo et al. (2007).
More than 10 stars were initially used to check variability and photometric errors.
Aperture photometry of comparison stars and quasars
was performed using the PHOT task in IRAF, with a 10 pixel (corresponding to 13'') aperture
and with an annulus between 15 and 20 pixels for sky subtraction, based on
our initial tests with images of various quality seeing. Photometric
errors of individual stars were typically less than 1\%.
We chose three to five unsaturated stars brighter than each quasar for differential
photometry. First, we calculated the magnitude difference between each comparison star
and the target quasar. Second, we normalized the magnitude difference 
to zero by subtracting the mean difference over all observed epochs.
Photometric errors of comparison stars were added in quadrature and then 
the photometric error of the target quasar was added on top of it,
yielding total errors of $\sim$1\% for relative flux changes.  

In Figure 1 we show relative B band flux variability. All five quasars showed
significant flux variability in the B band.
We find peak-to-peak variations in the range of 9\%-24\% and 
rms variations\footnote{We calculate rms variations using fluxes in logarithmic scale throughout the paper.} in the range of 3\%-8\%.
We note that since the MgII line flux is also included in the B band photometry, B band flux
variability cannot be considered as the continuum (such as 3000\AA) variability
unless the MgII line flux is constant (see \S3.4 for more details).
Nevertheless, the detected B band flux variations clearly demonstrate that our sample 
quasars have UV continuum variability.

\subsection{flux calibration using photometric measurements}

The flux uncertainty of the ground-based spectroscopy is typically larger than ~10\%
due to the slit losses, sky transparency, and various seeing conditions.
In reverberation mapping studies, non-varying narrow emission lines, 
such as [\ion{O}{3}] originated from far extended low-density
regions, are used as an internal flux calibrator. 
Thus, flux variations of broad lines can be measured by normalizing 
each spectrum to a constant flux of the narrow line (e.g. Peterson et al. 1998; 
Woo et al. 2007).
In the vicinity of the MgII line, there is no strong narrow line useful as an internal flux calibrator. The [OII] line at 3727\AA~is far from MgII in spectral range and much
weaker than MgII, not desirable as an internal calibrator.
However, since the MgII line is much stronger than H$\beta$, 
measuring its relative variation can be robustly done even without an internal flux calibrator as we demonstrate below. 

Instead of relying on the precision of flux calibration with spectrophotometric stars,
we normalize each single-epoch spectrum to match photometric measurements obtained at
the same night.
Using B band response function, we calculate synthetic magnitude from the observed spectrum
and then normalize the spectrum by multiplying a scale factor in order to match the B band magnitude 
measured from photometry. The amount of flux calibration with a scale factor was 
typically 10\%-20\% of the original flux.

We note that spectroscopy and photometry observations were not 
performed at the exact same time. Thus, spectroscopic flux calibration is uncertain within the level
of intra-night variations, which seem to be negligible for Seyfert 1 galaxies and quasars
based on the study by Webb \& Malkan (2000). Due to the limited telescope time for
photometry, we have not attempted to measure intra-night variations for individual objects,
which can be tested in the future.

\begin{figure}
\epsscale{1.0}
\plotone{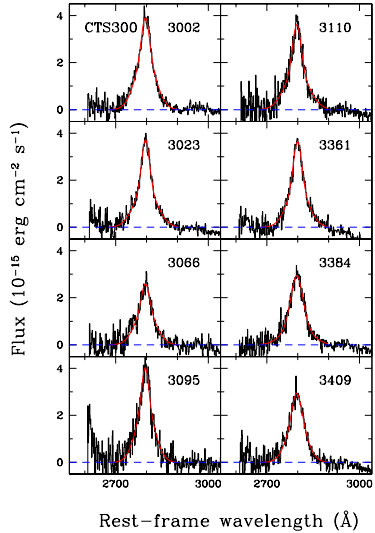}
%\plotone{fit300.eps}
\caption{A series of continuum subtracted spectra around the MgII line of CTS300.
The MgII lines are modeled with a multi-Gaussian profile (red).
Observing dates (JD-2450000) are denoted in the upper right conner
of each panel.
}
\label{fitCTS300}
\end{figure}

\begin{figure}
\epsscale{1.0}
\plotone{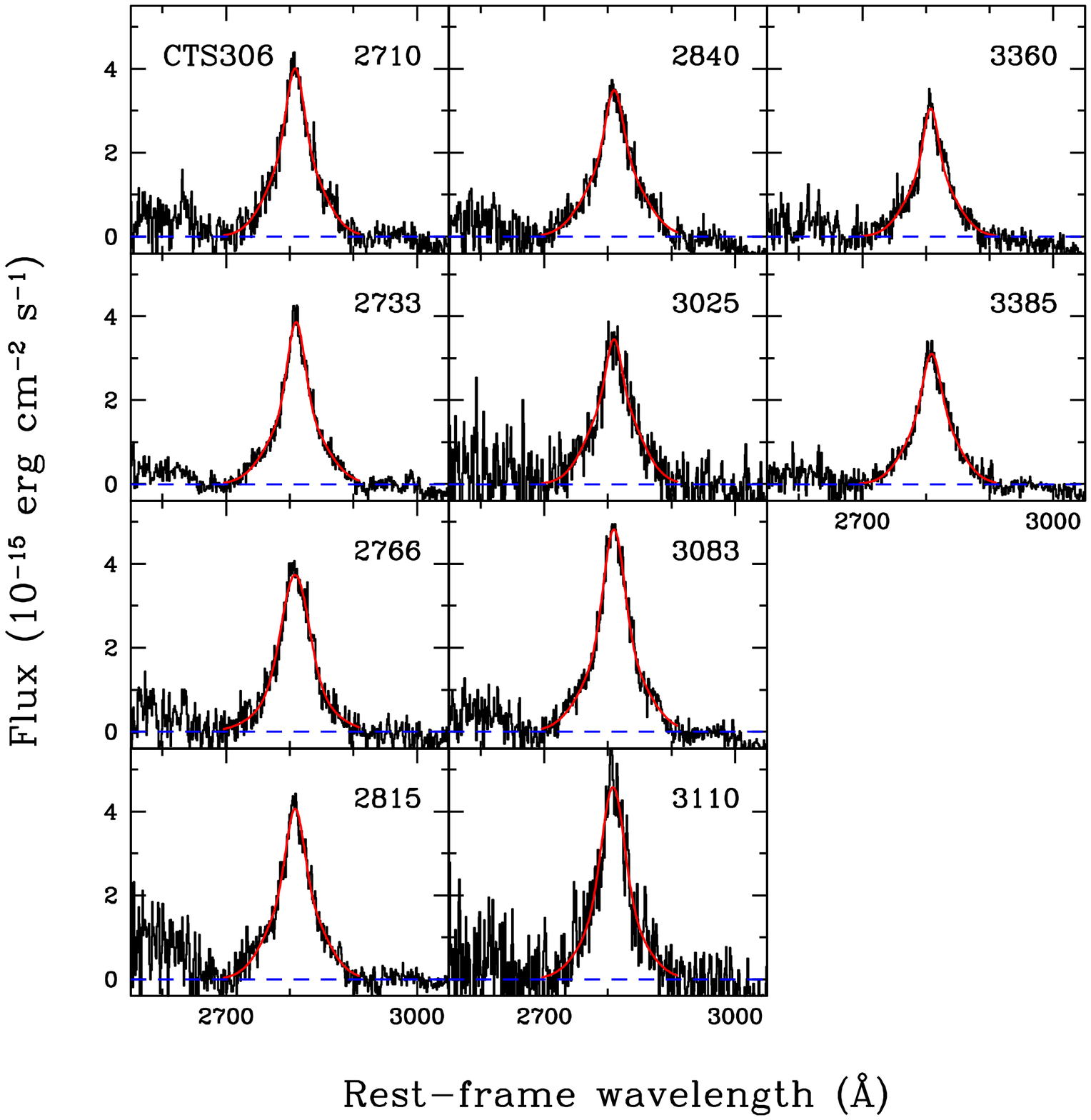}
%\plotone{fit306.eps}
\caption{As in Figure~\ref{fitCTS300} for CTS306.
}
\end{figure}

\begin{figure}
\epsscale{1.0}
\plotone{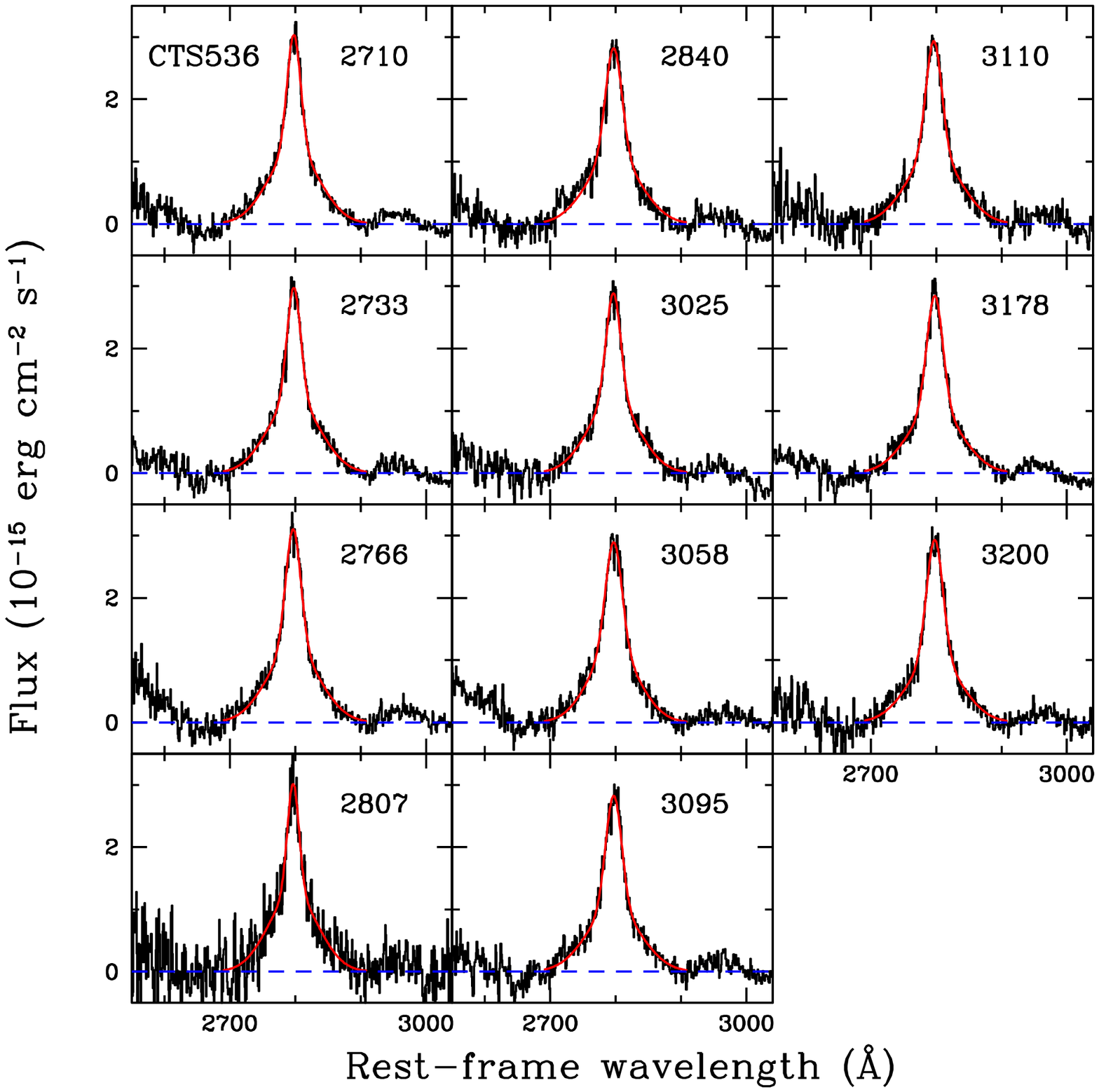}
%\plotone{fit536.eps}
\caption{As in Figure~\ref{fitCTS300} for CTS536.
}
\end{figure}

\begin{figure}
\epsscale{1.0}
\plotone{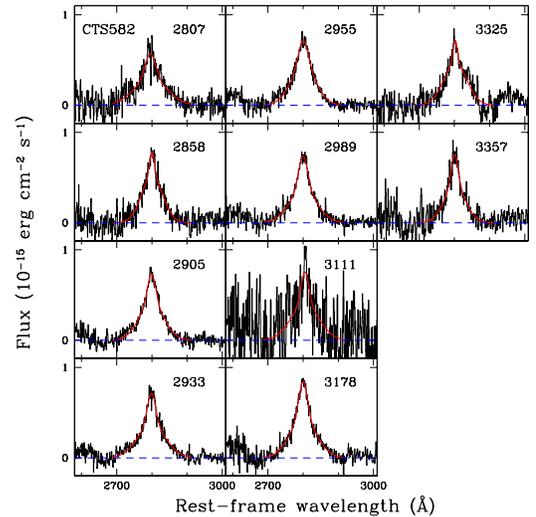}
%\plotone{fit582.eps}
\caption{As in Figure~\ref{fitCTS300} for CTS582.
}
\end{figure}

\begin{figure}
\epsscale{1.0}
%\plotone{fit667.eps}
\plotone{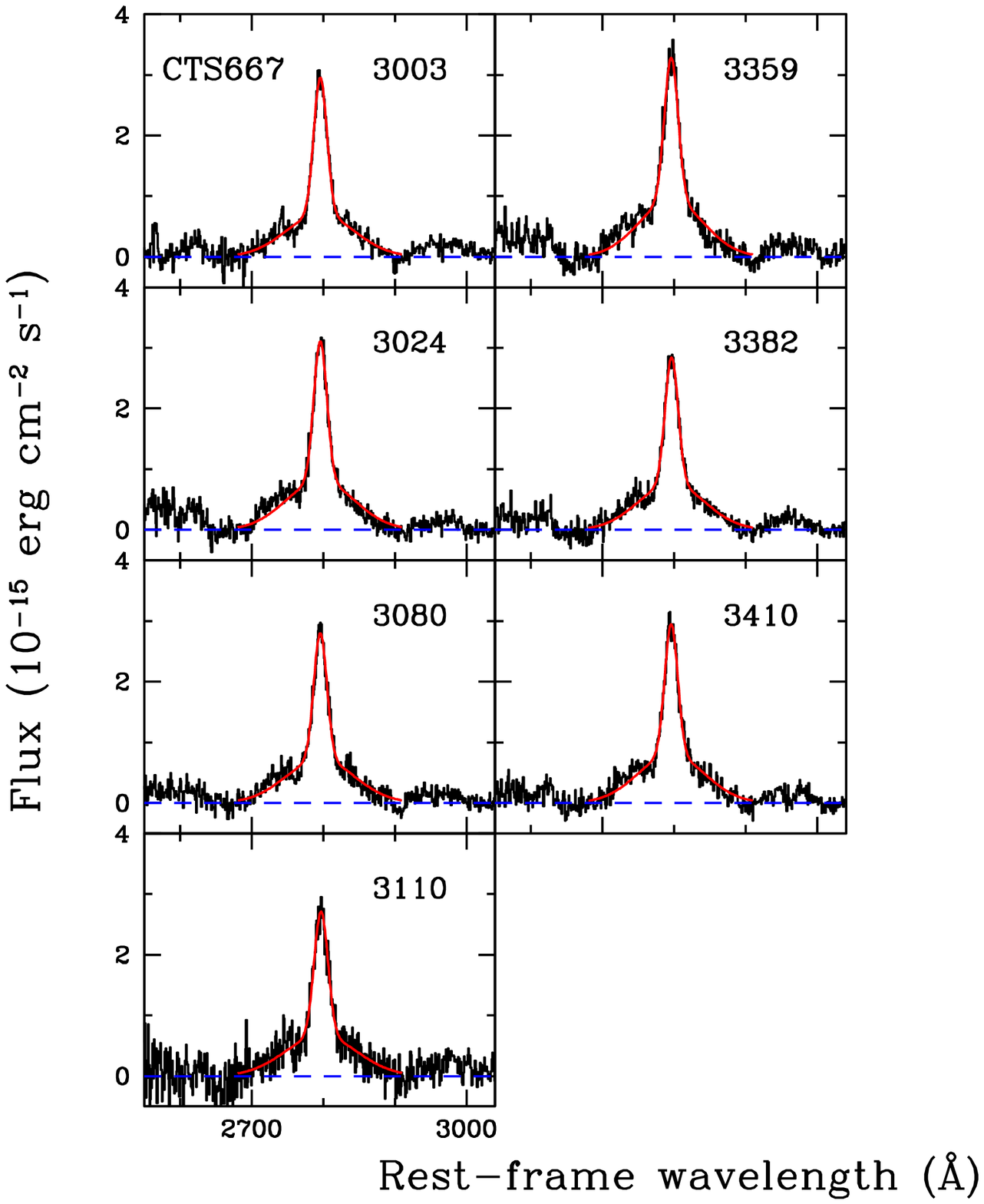}
\caption{As in Figure~\ref{fitCTS300} for CTS667.
}
\end{figure}
\subsection{MgII line and continuum flux measurements}

Using the flux calibrated spectra, we measure the flux and width of the MgII line
in several steps as similarly done in other studies (e.g. McGill et al. 2007).
First, we fit the continuum under MgII with a straight line 
using the average flux around $\sim$2700\AA~ and $\sim$2900\AA~ with 40\AA~
wide windows. We linearly interpolate these two continuum points to define
an underlying continuum. 
Second, we measure the MgII line flux by integrating the flux 
using the continuum subtracted spectrum. 
We note that defining a true continuum under the MgII line
is not straight forward because of the presence of broad FeII emission blends (small blue bump). Hence, depending on how the underlying continuum is defined, 
the measured MgII line flux can vary. 
Nevertheless, as far as we keep the same windows for determining the continuum,
we can measure the relative changes in the MgII line flux between two different epochs.
We did not subtract the broad FeII emission, which could contribute small amount of
flux to the wings of the MgII line. However, the relative flux changes in MgII will not be
significantly affected by the contribution of the FeII emission.

The random error on the MgII line flux measurements is typically 3\% level (Table 3). 
By adding $\sim$2\% systematic error of the flux calibration to the random error,
we take 5\% as an upper limit to the total error of the MgII line flux measurements.
A series of continuum-subtracted spectra for individual objects
are presented in Figures 2-6, where variation of the MgII line flux 
can be qualitatively detected.

From each spectrum the monochromatic continuum flux was measured at 3000\AA~ 
by averaging the flux within 40\AA~ window.
We note that the size of the window does not affect
the continuum variability. Typical errors on the continuum flux
measurements were 2\%-3\% depending on the S/N of each spectrum (see Table 3).

\begin{figure}
\epsscale{1.1}
\plotone{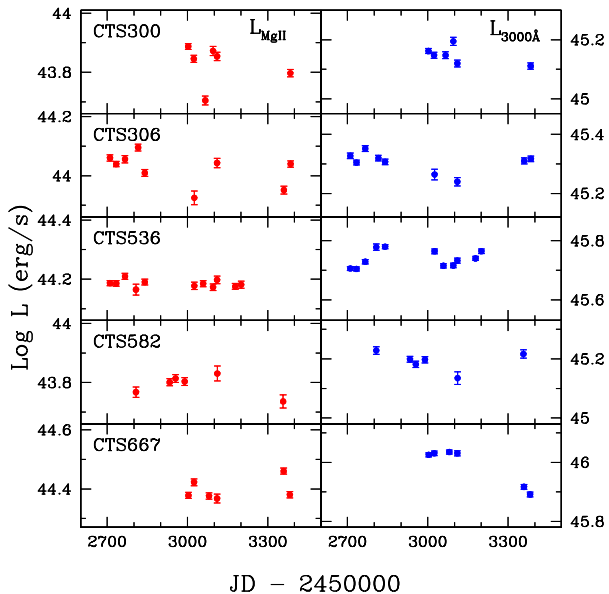}
%\plotone{vari.eps}
\caption{Variability of the MgII line flux (left panels) and the continuum flux at 3000\AA~(right panels) for the sample.
}
\end{figure}

\subsection{Flux variability of the MgII line and continuum}
\label{ssec:specvar}

In Figure 7 we present light curves of the MgII line flux (left) 
and the continuum flux at 3000\AA~ (right). All five quasars showed continuum (3000\AA)
flux variability, with peak-to-peak variations of 17\%-38\% and 
rms variations of 6\%-15\%.
% for the continuum luminosity at 3000\AA.
Continuum flux variation shows consistent pattern as B band flux variation
in the case of CTS536, for which the MgII line flux has been almost constant during the 
monitoring period. 
However, for the other objects, the continuum flux variations are different from B band flux
variations depending on the relative contribution of the MgII line flux.

In the case of MgII line flux, we detected 23\%-50\% peak-to-peak changes
and  8\%-17\% rms variations for four objects. In contrast, one object, CTS536, did not show
clear variation within the measurement error 5\%. 
The MgII line flux variation of these four quasars is similar to that of 
local Seyfert galaxies and high redshift quasars (e.g. Clavel et al. 1991; Trevese et al. 2007), 
implying that intermediate redshift quasars have similar characteristics of the MgII line flux variability,
% and that of the CIV line of in the paper by Kaspi et al. (2007)
although the actual amplitude is highly variable depending on individual objects
and the monitoring time base-line.

%In Figure 8 we present two mean spectra, dividing all epochs into two groups, high and low line flux states,
%to show relative change in line flux. The MgII line profiles of two mean spectra show no clear changes.
%Most of flux changes are in the core and flux changes in the wing is not dominant,
%implying that the contamination from the broad FeII blends below the MgII line is not significant.

\subsection{Line width variability}

To measure the MgII line width, we fit the line profile with multi-Gaussian models and 
determine the best fit model based on $\chi^{2}$ statistics.
Fitting the line profile is necessary
for low quality spectra, for which FWHM can not be easily defined on the data. 
We measured FWHM on the best fit model as demonstrated in Figures 2-6 (red lines). 
In the case of CTS667, FeII contamination is present in the wings of the MgII line profile and our multi-Gaussian model also fits the broad FeII component. 
However, FWHM of the MgII line is not significantly affected by the shape of the wings.
Thus, we simply ignored the FeII component although this will increase the uncertainties
of MgII line width measurements, which is dominated by systematic errors from continuum 
subtraction and assumed to be $\sim$5\%.

In Figure 8 we compared measured FWHM with the continuum luminosity
for each object.
The rms variation of the MgII line width, ranging from 4\% to 15\% (see Table 3), is
similar to the rms variation of continuum flux, suggesting 
a photoionization origin of the MgII line.
The reason why we do not clearly see a correlation between the line width and continuum 
luminosity, as expected from a photoionization model (e.g.
FWHM $\propto$ L$_{3000\AA}^{1/2}$), is probably due to the combined effects of measurement and
systematic errors on the line width and relatively small flux variations of the continuum.

%We measured the FWHM of MgII to investigate line width variability.
%Using multi-Gaussian profiles, we fit the profile of MgII and measured FWHM on the fit.
%In Figure 4, we demonstrated continuum subtracted spectra with multi-Gaussian fit.
%A combination of two Gaussian profiles fits the MgII line relatively well.

\begin{deluxetable*}{lcrrrrrrrrrrr}
%\rotate{}
\tablewidth{0pt}
\tablecaption{Variability measurements}
\tablehead{
\colhead{Target}         &
\colhead{N}  &
\colhead{N}  &
\colhead{MgII}     &
\colhead{rms}     &
\colhead{error}     &
\colhead{L$_{3000}$}     &
\colhead{rms}     & 
\colhead{error}    & 
\colhead{F}    &  
\colhead{rms}    &  
\colhead{F}    &  
\colhead{L$_{3000}$}    \\ 
\colhead{(1)} &
\colhead{(2)} &
\colhead{(3)} &
\colhead{(4)} &
\colhead{(5)} &
\colhead{(6)} &
\colhead{(7)} &
\colhead{(8)} &
\colhead{(9)} &
\colhead{(10)} &
\colhead{(11)} &
\colhead{(12)} &
\colhead{(13)} }
\tablecolumns{13}
\startdata 
CTS300 &8  & 6  & .17 (50\%)   & .067 (17\%) & .013 (3\%) & .08 (20\%) & .028 (7\%) & .011 (3\%) & 5051 & .035 (8\%)  & 5197 & 45.15  \\
CTS306 &10 & 9  & .16 (45\%)   & .054 (13\%) & .013 (3\%) & .12 (32\%) & .032 (8\%) & .011 (3\%) & 5494 & .035 (8\%)  & 5540 & 45.30  \\
CTS536 &11 & 11 & .04 (10\%)   & .012 (3\%)  & .011 (3\%) & .07 (17\%) & .027 (6\%) & .008 (2\%) & 4020 & .033 (8\%)  & 4101 & 45.74  \\
CTS582 &10 & 6  & .13 (35\%)   & .034 (8\%)  & .017 (4\%) & .09 (23\%) & .030 (7\%) & .013 (3\%) & 5102 & .063 (15\%) & 5144 & 45.19  \\
CTS667 &7  & 6  & .09 (23\%)   & .036 (9\%)  & .012 (3\%) & .14 (38\%) & .060 (15\%)& .008 (2\%) & 2907 & .015 (4\%)  & 2862 & 45.99  \\
\enddata
\tablecomments{
Col. (1): Target ID.  
Col. (2): Number of observed epochs for spectroscopy.
Col. (3): Number of epochs with flux calibration using available photometry.
Col. (4): peak to peak variation of MgII line flux in logarithmic scale.
Col. (5): rms variation of MgII line flux in logarithmic scale.
Col. (6): mean error on MgII line flux measurement in logarithmic scale.
Col. (7): peak to peak variation of continuum luminosity at 3000\AA~ in logarithmic scale.
Col. (8): rms variation of continuum luminosity at 3000\AA~ in logarithmic scale.
Col. (9): mean error on continuum luminosity measurement in logarithmic scale.
Col. (10): mean FWHM of the MgII line in km s$^{-1}$.
Col. (11): rms variation of FWHM in logarithmic scale.
Col. (12): FWHM of the MgII line in km s$^{-1}$, measured from mean spectra.
Col. (13): L$_{3000\AA}$ in erg s$^{-1}$, measured from mean spectra.
}
\end{deluxetable*}

\begin{figure}
\epsscale{1.2}
\plotone{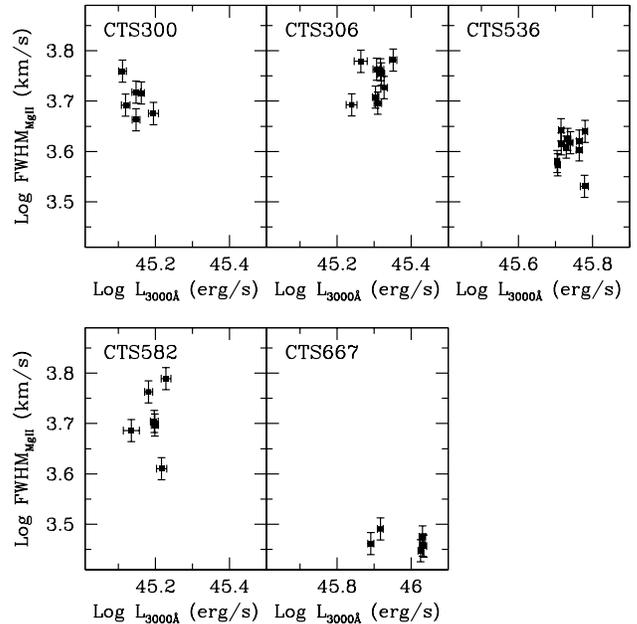}
\caption{MgII line width variability. 5\% error was assumed for the FWHM measurements.
}
\label{time}      
\end{figure}
\subsection{Black hole mass estimates}

We estimated black hole mass using the virial technique
based on the empirical size-luminosity relation of the reverberation sample
(Kaspi et al. 2005; Bentz et al. 2006).
Combining the MgII line FWHM and the continuum luminosity at 3000\AA,
measured from the mean spectra of each object, 
black hole mass was determined with the empirical formula from McGill et al. (2007):

\begin{equation}
M_{BH} = 6.77 \left(L_{\rm 3000} \over 10^{44} \mbox{ \rm erg} s^{-1} \right)^{0.47} \left(\rm FWHM_{\rm MgII} \over 1000 \mbox{ \rm km s}^{-1} \right)^{2} M_{\odot}. 
\label{eq:MK}
\end{equation}

The estimated black hole mass of the sample quasars ranges between 10$^{8.6}$ and 10$^{8.9}$ M$_{\odot}$ (see Table 1). 
By multiplying a factor of 5.9 to the continuum luminosity at 3000\AA~(McLure \& Dunlop 2004), 
we estimated bolometric luminosity and computed the Eddington ratio. The determined
Eddington ratio ranges from 0.1 to 1.1 as tabulated in Table 1, 
indicating our sample quasars are highly accreting luminous quasars. 

\section{Discussion and Conclusions}

We monitored and measured variability of 5 high-luminosity quasars 
at $0.4\lesssim z \lesssim 0.6$. From the photometry, we detected 
10\%-24\% peak-to-peak flux variations and 3\%-8\% rms variations 
in the B band which corresponds to the rest-frame 2500\AA-3600\AA~range. 
After spectroscopic flux calibration using photometric measurements,
we measured the MgII line flux and the continuum flux at 3000\AA~from the
observed spectra. MgII flux variations were detected for 4 objects with 
23\%-50\% peak-to-peak variations and  8\%-17\% rms variations.
One object did not show clear variations 
in the MgII line flux within the measurement error ($\sim$5\%) over the two year monitoring
period. By measuring FWHM of the MgII line, we measured variability of line width for the
same quasars, which showed 4\%-15\% rms variations.

These variability detections are consistent with a picture that the MgII line 
is variable as other broad lines, such as Balmer lines or high ionization lines 
(Ly$\alpha$ and CIV). The MgII line of these intermediate redshift quasars 
probably originates in the same region as H$\beta$, as shown in the cases of 
NGC 3783 and NGC 4151, for which similar time-lags were measured from 
H$\beta$ and MgII lines (Reichert et al. 1994; Peterson et al. 2004; Metzroth et al. 2006).
It is possible that for some particular objects MgII line emitting region is 
more extended than H$\beta$ line emitting region, 
and hence MgII line variability may not be detected in short time scales.
This scenario was suggested for NGC 3516, which showed no MgII line flux variations 
(Goad et al. 1999). However, it is not clear whether NGC 3516 is an abnormal object 
with an extended MgII line region, or substantial fraction of quasars have a similarly extended MgII line region.
%Clear understanding of physics of MgII broad line region is yet to be done.
%Whether MgII line is originated from the same region as H$_\beta$ line can be addressed 
Monitoring flux variations of the MgII line together with the H$\beta$ line 
using a larger sample of AGNs is necessary to properly address these issues.

Although we did not attempt to do reverberation analysis
due to the short time baseline with insufficient numbers of epochs, 
it is shown that the MgII line can be used for reverberation mapping 
for even luminous quasars at intermediate redshifts.
Compared to the local Seyfert galaxies, 
a longer-term monitoring program is necessary for these objects
due to the 1+z effect on the observed time scale 
as well as larger intrinsic reverberation time scales for more luminous quasars.
The estimated time-lags for the sample quasars, based on the continuum 
luminosity at 3000\AA~ and the size-luminosity relation (McLure \& Dunlop 2004),
range from 74 to 217 light-days. Thus, a few year monitoring program with 
synchronous photometry and spectroscopy can
reveal the reverberation time scales of the MgII line for the sample.
Small telescopes (1-3m) can be very useful tools for MgII line 
reverberation mapping since MgII is much stronger than H$\beta$ which
has been generally used in reverberation mapping of local AGNs. 
Regular monitoring and service observation is a key for 
a success in future reverberation mass measurements.

\medskip

\acknowledgments

This work is based on data collected at the CTIO through the SMARTS consortium. 
I am grateful to Meg Urry for support and the inspiration for the project during my stay 
at the Yale Center for Astronomy and Astrophysics.
I thank Tommaso Treu and Charles Bailyn for useful discussions and suggestions
and the referee for useful suggestions.
I thank all CTIO staffs for service observations, Rebecca Winnick for scheduling,
and Brendan Cohen for data collection and initial reduction works. 
I specially thank Suzanne Tourtellotte for data archiving and handling.

%\clearpage


\begin{thebibliography}{}
\bibitem[Bentz et al.(2006)]{2006ApJ...644..133B} Bentz, M.~C., Peterson, B.~M., Pogge, R.~W., Vestergaard, M., \& Onken, C.~A.\ 2006, \apj, 644, 133
%\bibitem[Bentz et al.(2006)]{2006ApJ...651..775B} Bentz, M.~C., et al.\ 2006b, \apj, 651, 775 
\bibitem[Bonning et al.(2007)]{2007ApJ...659..211B} Bonning, E.~W., Cheng, 
L., Shields, G.~A., Salviander, S., \& Gebhardt, K.\ 2007, \apj, 659, 211 
\bibitem[Blandford \& McKee(1982)]{1982ApJ...255..419B} Blandford, R.~D., \& McKee, C.~F.\ 1982, \apj, 255, 419 
\bibitem[Clavel et al.(1991)]{1991ApJ...366...64C} Clavel, J., et al.\ 
1991, \apj, 366, 64 
\bibitem[Croton et al.(2006)]{2006MNRAS.365...11C} Croton, D.~J., et al.\ 
2006, \mnras, 365, 11
\bibitem[Di Matteo et al.(2005)]{2005Natur.433..604D} Di Matteo, T., Springel, V., \& Hernquist, L.\ 2005, \nat, 433, 604
\bibitem[Davis et al.(2007)]{2007arXiv0707.1456D} Davis, S.~W., Woo, J.-H., \& Blaes, O.~M.\ 2007, ApJ, in press (astroph/07071456)
\bibitem[Goad et al.(1999)]{1999ApJ...524..707G} Goad, M.~R., Koratkar, A.~P., Kim-Quijano, J., Korista, K.~T., O'Brien, P.~T., \& Axon, D.~J.\ 1999, \apj, 524, 707

\bibitem[Kaspi et al.(2000)]{2000ApJ...533..631K} Kaspi, S., Smith, P.~S., Netzer, H., Maoz, D., Jannuzi, B.~T., \& Giveon, U.\ 2000, \apj, 533, 631 
\bibitem[Kaspi et al.(2005)]{2005ApJ...629...61K} Kaspi, S., Maoz, D., Netzer, H., Peterson, B.~M., Vestergaard, M., \& Jannuzi, B.~T.\ 2005, \apj, 629, 61 
\bibitem[Kaspi et al.(2007)]{2007ApJ...659..997K} Kaspi, S., Brandt, W.~N., Maoz, D., Netzer, H., Schneider, D.~P., \& Shemmer, O.\ 2007, \apj, 659, 997
\bibitem[Maza et al.(1993)]{1993RMxAA..25...51M} Maza, J., Ruiz, M.~T., Gonzalez, L.~E., Wischnjewsky, M., \& Antezana, R.\ 1993, Revista Mexicana de Astronomia y Astrofisica, 25, 51 
\bibitem[Maza et al.(1995)]{1995RMxAA..31..159M} Maza, J., Ortiz, P.~F., Wischnjewsky, M., Antezana, R., \& Gonzalez, L.~E.\ 1995, Revista Mexicana de Astronomia y Astrofisica, 31, 159
\bibitem[Maza et al.(1996)]{1996RMxAA..32...35M} Maza, J., Wischnjewsky, M., \& Antezana, R.\ 1996, Revista Mexicana de Astronomia y Astrofisica, 32, 35
\bibitem[McGill et al.(2007)]{2007ApJ.000.0000M} McGill, , K. L., Woo, J.-H., Treu, T., \& Malkan, M. A. 2007, ApJ, 673, 703 
\bibitem[McLure \& Dunlop(2004)]{2004MNRAS.352.1390M} McLure, R.~J., \& Dunlop, J.~S.\ 2004, \mnras, 352, 1390
\bibitem[Metzroth et al.(2006)]{2006ApJ...647..901M} Metzroth, K.~G., 
Onken, C.~A., \& Peterson, B.~M.\ 2006, \apj, 647, 901
\bibitem[{{Netzer} \& {Trakhtenbrot}(2007)}]{N+T07} {Netzer}, H. \& {Trakhtenbrot}, B. 2007, \apj, 654, 754
%\bibitem[Onken et al.(2004)]{2004ApJ...615..645O} Onken, C.~A., Ferrarese, L., Merritt, D., Peterson, B.~M., Pogge, R.~W., Vestergaard, M., \& Wandel, A.\ 2004, \apj, 615, 645
%\bibitem[Peng et al.(2006)]{2006ApJ...649..616P} Peng, C.~Y., Impey, C.~D., 
%Rix, H.-W., Kochanek, C.~S., Keeton, C.~R., Falco, E.~E., Leh{\'a}r, J., \& 
%McLeod, B.~A.\ 2006, \apj, 649, 616 
%\bibitem[Peterson(1993)]{1993PASP..105..247P} Peterson, B.~M.\ 1993, \pasp, 105, 247 
\bibitem[Peterson et al.(1998)]{1998ApJ...501...82P} Peterson, B.~M., 
Wanders, I., Bertram, R., Hunley, J.~F., Pogge, R.~W., \& Wagner, R.~M.\ 1998, \apj, 501, 82
%\bibitem[Peterson \& Wandel(2000)]{2000ApJ...540L..13P} Peterson, B.~M., \& Wandel, A.\ 2000, \apjl, 540, L13 
%\bibitem[Peterson et al.(2002)]{2002ApJ...581..197P} Peterson, B.~M., et al.\ 2002, \apj, 581, 197
\bibitem[Peterson et al.(2004)]{2004ApJ...613..682P} Peterson, B.~M., et al.\ 2004, \apj, 613, 682 
%\bibitem[Peterson et al.(2005)]{2005ApJ...632..799P} Peterson, B.~M., et al.\ 2005, \apj, 632, 799
\bibitem[Reichert et al.(1994)]{1994ApJ...425..582R} Reichert, G.~A., et al.\ 1994, \apj, 425, 582 
\bibitem[Trevese et al.(2007)]{2007A&A...470..491T} Trevese, D., Paris, D., Stirpe, G.~M., Vagnetti, F., \& Zitelli, V.\ 2007, \aap, 470, 491
\bibitem[Treu et al.(2007)]{2007ApJ...667..117T} Treu, T., Woo, J.-H., 
Malkan, M.~A., \& Blandford, R.~D.\ 2007, \apj, 667, 117
\bibitem[Vestergaard(2002)]{2002ApJ...571..733V} Vestergaard, M.\ 2002, \apj, 571, 733
\bibitem[Vestergaard(2004)]{2004ApJ...601..676V} Vestergaard, M.\ 2004, \apj, 601, 676
%\bibitem[Vestergaard \& Peterson(2006)]{2006ApJ...641..689V} Vestergaard, M., \& Peterson, B.~M.\ 2006, \apj, 641, 689 
\bibitem[Wandel et al.(1999)]{1999ApJ...526..579W} Wandel, A., Peterson, B.~M., \& Malkan, M.~A.\ 1999, \apj, 526, 579 
\bibitem[Webmalkan]{web} Webb, \& Malkan, A. 2000, ApJ, 540, 652
\apj, 601, 676
\bibitem[Woo \& Urry(2002)]{2002ApJ...579..530W} Woo, J.-H., \& Urry, C.~M.\ 2002, \apj, 579, 530 
\bibitem[Woo et al.(2006)]{2006ApJ...645..900W} Woo, J.-H., Treu, T., Malkan, M.~A., \& Blandford, R.~D.\ 2006, \apj, 645, 900
\bibitem[Woo et al.(2007)]{2007ApJ...661...60W} Woo, J.-H., Treu, T., Malkan, M.~A., Ferry, M.~A., \& Misch, T.\ 2007, \apj, 661, 60
\end{thebibliography}
\end{document}